\documentclass[aps,pra,reprint,amsmath,amssymb]{revtex4-1}
\usepackage{braket}
\usepackage{graphicx}
\usepackage[colorinlistoftodos]{todonotes}

\begin{document}

\title{One- and two-photon scattering from generalized $V$-type atoms}

\author{Eduardo S{\'a}nchez-Burillo}
\affiliation{Instituto de Ciencia de Materiales de Arag{\'o}n y Departamento de F{\'\i}sica de la Materia Condensada, CSIC-Universidad de Zaragoza, Zaragoza, E-50009, Spain}
\author{David Zueco}
\affiliation{Instituto de Ciencia de Materiales de Arag{\'o}n y Departamento de F{\'\i}sica de la Materia Condensada, CSIC-Universidad de Zaragoza, Zaragoza, E-50009, Spain}
\affiliation{Fundacion ARAID, Paseo Mar{\'\i}a Agust{\'\i}n 36, Zaragoza 50004, Spain}
\author{Luis Mart{\'\i}n-Moreno}
\affiliation{Instituto de Ciencia de Materiales de Arag{\'o}n y Departamento de F{\'\i}sica de la Materia Condensada, CSIC-Universidad de Zaragoza, Zaragoza, E-50009, Spain}
\author{Juan Jos{\'e} Garc{\'\i}a-Ripoll}
\affiliation{Instituto de F{\'\i}sica Fundamental, IFF-CSIC, Calle Serrano 113b, Madrid 28006, Spain}

\begin{abstract}
The one- and two-photon scattering matrix $S$ is obtained analytically for a one-dimensional waveguide and a point-like scatterer with $N$ excited levels (generalized $V$-type atom).
We argue that the two-photon scattering matrix contains sufficient information to distinguish between different level structures which are equivalent for single-photon scattering, such as a $V$-atom with $N=2$ excited levels and two two-level systems. In particular, we show that the scattering with the $V$-type atom exhibits a destructive interference effect leading to two-photon Coupled-Resonator-Induced Transparency, where the nonlinear part of the two-photon scattering matrix vanishes when each incident photon fulfills a single-photon condition for transparency. 
\end{abstract}

\maketitle

\section{Introduction}

The theoretical study of the scattering of photons by isolated few-level systems is now an essential tool for describing transport experiments using photons interacting with systems like quantum dots or atoms in photonic crystals\ \cite{sollner14,lodahl15review,Goban2015}, superconducting qubits in open transmission lines\ \cite{astafiev10,hoi11,haeberlein15} or atoms in dielectric waveguides \cite{Vetsch2010}. The challenges and possibilities offered by experiments with multiphoton wavepackets have motivated the development of new techniques for solving the dynamics associated to strong light-matter interaction. Consequently, there has been a significant progress from initial works based on few-photon wavefunctions \cite{shen05a,shen05b}, going from real space calculations \cite{Baranger2010,Roy2013,Ballestero2014}, Green function based techniques \cite{Baranger2013a,Laakso2014} or input-output theory \cite{fan10} to field-theoretical methods \cite{shi15,xu15}, as well as numerical approaches \cite{Longo2009,Longo2010,Longo2011,Sanchez-Burillo2014,Sanchez-Burillo2015,Kocabas2016}. These techniques open the door to the study of multi-photon processes and nonlinear phenomena in many-qubit systems, the properties of collectively emitted and non-classical states of light, or the engineering of photon-mediated interactions and collective dissipative dynamics.

In this work we study the scattering properties of one and two photons traveling in a 1D waveguide and impinging on a multilevel quantum system. In particular, we focus on a generalized $V$-level scheme, consisting on a single ground state that can be excited to $N$ different states which are uncoupled among them [cf. Fig.\ \ref{fig:setup}a], which we will denote as $V^{(N)}$-atom. The case $N=1$ describes a two-level system (2LS), and the case $N=2$ describes a $V$-atom (which can be either an actual atom or an effective one, e.g., made with inductively coupled transmons\ \cite{Dumur15}). Beyond these cases, the $V^{(N)}$-level structure describes many atomic spectra. For instance, the ground state $\ket{0}$ can represent one hyperfine state whose excitation is constrained, due to different selection rules, to a subset of atomic states $\{\ket{i}\}_{i=1}^N$ depending on the polarization properties of the incoming light.  Also, a $V^{(N)}$-atom can describe $N$ different two-level systems influenced by a blockade mechanism that prevents the simultaneous excitation of two or more absorbers [cf. Fig.\ \ref{fig:setup}b], a feature characteristic of Rydberg atoms used in various quantum information and quantum simulation tasks\ \cite{Jaksch2000,Lukin2001,Saffman10}.

We also compare the scattering properties in the $N=2$ case with those for two independent 2LS. The scattering of a single photon by a $V$-atom is the same as by two collocated 2LS. In particular, in both situations, the single-photon scattering presents the so-called Coupled-Resonator-Induced Transparency (CRIT).  In this phenomenon, akin to Electromagnetically Induced Transparency (EIT) \cite{Harris1997}, perfect photon transmission occurs due to Fano-type interference between virtual transitions to the coupled levels in the resonators \cite{Smith2004}.  However, we show that there are significant differences between the two-photon resonance fluorescence arising from scattering by a $V^{(N)}$-atom and that from scattering by two collocated 2LS.  For instance,  scattering by a $V^{(N)}$-atom presents two-photon CRIT, while that by the collocated 2LS does not. 

The structure of this paper is as follows. In Sect.\ \ref{sec:model} we introduce the Hamiltonian for photons propagating in a one-dimensional waveguide interacting with a $V^{(N)}$-atom. In Sect.\ \ref{sec:scattering} we develop the single-photon and two-photon scattering theory for this model, using the input-output formalism. Sect.\ \ref{sec:discussion} applies our results to a number of idealized experiments. In Subsect.\ \ref{sec:2ph-fluorescence} we compare the single and two-photon scattering by a $V$-atom with that by two 2LS. We show that only the two-photon spectrum distinguishes between both cases. Subsect.\ \ref{sec:2photon-eit} takes this idea further and demonstrates that the two-photon scattering spectrum by a $V^{(N)}$-atom presents instances of perfect transmission and no-nonlinearity.  These situations arise from a destructive interference phenomenon that mimics that of single-photon CRIT. 
\section{Model and input-output theory}
\label{sec:model}

\begin{figure}
\includegraphics[width=0.9\linewidth]{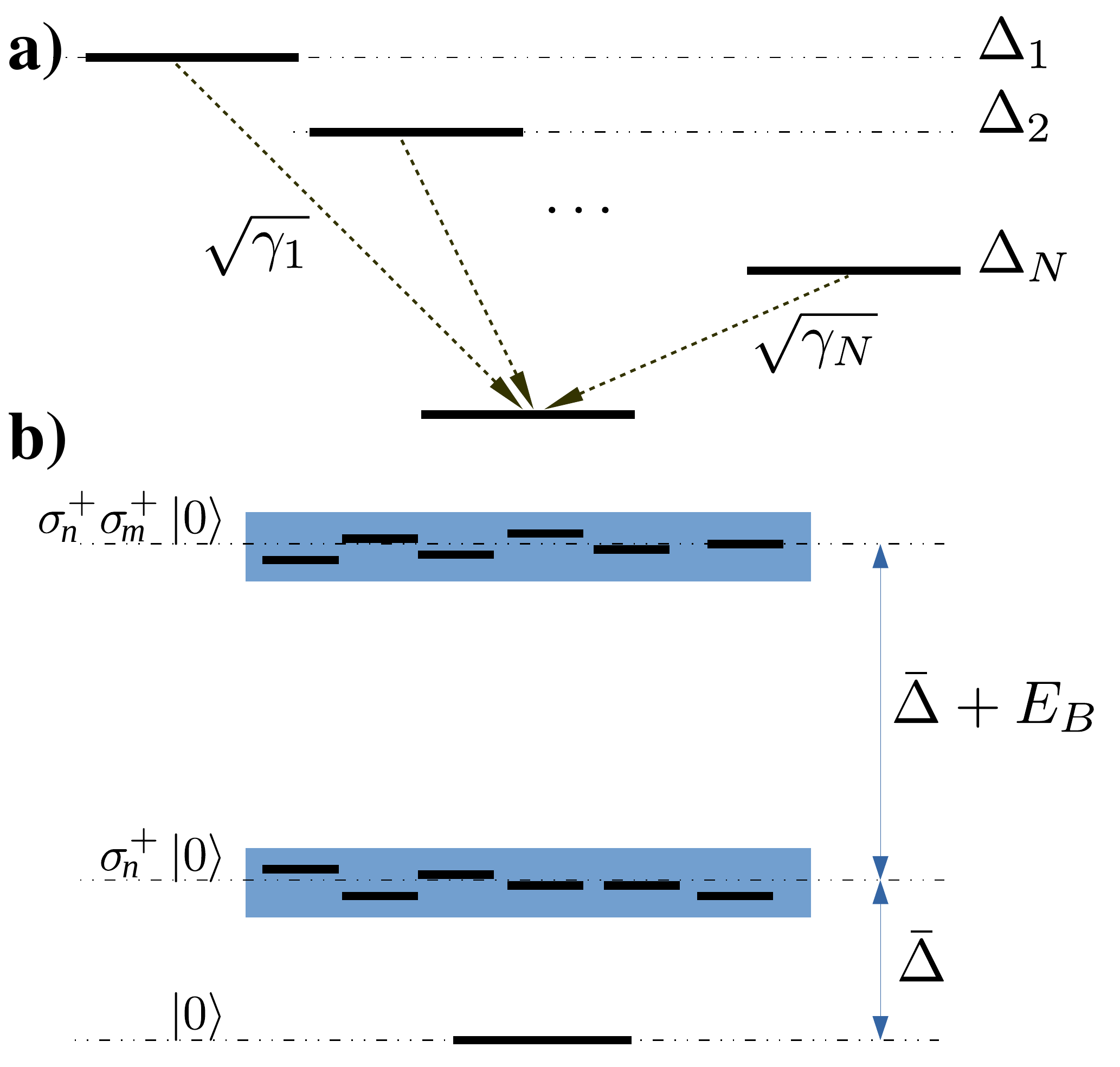}
\caption{(a) $V^{(N)}$-atom. We study a point-like particle interacting with a continuum of propagating modes. The quantum impurity has $N$ excited levels with energies $\Delta_i,\,i=1\ldots N$ and decay rates $\gamma_i$, which we use to parameterize the light-matter interaction. (b) The previous level structure can be a good approximation for $N$ two-level systems presenting a blockade mechanism \cite{Jaksch2000,Lukin2001,Saffman10}, where the excited states $\sigma^+_i\ket{0}$ of the respective atoms or qubits have a strong repulsive interaction, $E_B\gg \gamma_i$, thus preventing simultaneous multiple excitations. }
\label{fig:setup}
\end{figure}

Our model considers photons propagating in a one-dimensional waveguide, interacting with a point-like scatterer characterized by $N+1$ discrete quantum levels ($N$ excited levels and the ground state) [cf. Fig.\ \ref{fig:setup}].  This situation is an extension to the $N=1$ case considered in \cite{fan10}. Following that work, we use two common approximations. First, we linearize the dispersion relation of photons around the energy of the incoming photons $\omega_0$, $\omega(k) \simeq \omega_0 + v_g |k\mp k_0|$ for right- and left-moving photons respectively. Here, $k_0$ is the momentum such that $\omega(\pm k_0)=\omega_0$ and $v_g$ is the group velocity at $k=\pm k_0$. We will set the zero of energies at $\omega_0$. In addition, we will refer our momentum to the reference momentum $\pm k_0$ for right- and left-moving photons respectively. Then, we can rewrite the dispersion relation as $\omega(k)=v_g|k|$. Secondly, the interaction (dipole) Hamiltonian between the photon and the scatterer is treated within the Rotating-Wave-Approximation (RWA), which preserves the number of excitations. These approximations are excellent when the photon frequency is far from a band edge and the coupling strength is much smaller than the excitation energy. 

The Hamiltonian then reads ($\hbar=1$)
\begin{align}\label{H}
H &= \sum_{n=1}^N \Delta_n\ket{n}\bra{n} +  \sum_{s\in\pm}
\int_{-\infty}^\infty \omega \, a_{s\omega}^\dagger a_{s\omega} \, \mathrm{d}\omega \\
&+ \sum_{s=\pm}\sum_{n=1}^N \frac{g_{sn}}{\sqrt{v_g}} \int_{-\infty}^\infty  (\sigma^+_n a_{s\omega} + \sigma^-_n a^\dagger_{s\omega} ) \, \mathrm{d}\omega.\nonumber
\end{align}
Here $\sigma^+_n = \ket{n}\bra{0}$ and $\sigma^-_n = \ket{0}\bra{n}$ are ladder operators for the generalized $V-$ atom, $s\in\{\pm\}$ represents the two directions of propagation of the photons and $a_{s\omega}$ is the bosonic annihilation operator for a photon with energy $\omega$ and direction $s$. The excitation energies are denoted by $\Delta_n$ and $g_{sn}$ are the coupling strengths of the corresponding transitions. Notice that the integration range has been extended from $-\infty$ to $+\infty$, which is valid if the energies of the incident photons are close enough to the linearization point $\omega_0$ \cite{Loudon2000}. From now on, we will assume the integrals go always from $-\infty$ to $\infty$ and we will drop the integration limits.

Notice that this Hamiltonian contemplates the possibility of dissimilar couplings from the emitter to left-moving and right-moving photons. This is interesting in its own right, as the waveguide could be chiral and allow the propagation in only one direction. It is also interesting as a theoretical device, as the scattering properties in the non-chiral case ($g_{+n}=g_{-n}$) can be related to those of the chiral one ($g_{-n}=0$, $g_{+n}=g_n$)\cite{fan10}, which are easier to compute because the latter involves a single branch of photons. We will follow this  approach, performing first the calculations for a chiral waveguide and explicitly providing the results for the non-chiral case later on. As we will just consider one kind of photon, we will have just one set of bosonic operators for the chiral computations, $a_\omega$. Besides, if we take length units such that $v_g=1$, the dispersion relation is $\omega(k)=k$. Therefore, we can use either $\omega$ or $k$ without distinction. Following \cite{fan10}, we write all the expressions in terms of $k$.

The Heisenberg equations for the atom and photon operators with the chiral model read:
\begin{align}
i\partial_t a_k(t) &= k a_k (t)+ \sum_{n=1}^N g_n \sigma^-_n(t),\\
i\partial_t \sigma^-_n (t)&= \Delta_n \sigma^-_n(t) +
\sum_{m=1}^N \int g_m \, c_{mn}(t)a_k(t)\mathrm{d}k,
\end{align}
where the operators $c_{mn}:=\delta_{mn}\ket{0}\bra{0} - \sigma^+_m\sigma^-_n$.

In order to extract the scattering properties, the in-out formalism introduces the asymptotic free fields $a_\text{in}(t):= 1/\sqrt{2 \pi} \int_{0}^{\infty} {\rm d}k a_k (t_0){\rm e}^{-i k (t-t_0)}$ and  $a_\text{out}(t):= 1/\sqrt{2 \pi} \int_{0}^{\infty} {\rm d}k a_k (t_1) {\rm e}^{-i k (t-t_1)}$,  where $t_0 \to -\infty$ and $t_1 \to \infty$ \cite{Gardiner1985}.
Following the derivations in \cite{fan10} for the case of a 2LS, mutatis mutandis, the ``out'' fields in the case of general $N$ are related to the ``in'' fields through the time evolution of the ladder operators 
\begin{equation}\label{eq:in_out}
a_\text{out}(t) = a_\text{in}(t) -i \sum_{n=1}^N \sqrt{2\gamma_n} \, \sigma^-_n(t),
\end{equation}
where $\gamma_n = \pi g_n^2$ is the spontaneous emission rate of the $n$-th transition ($\ket{n}\to\ket{0}$) coupled to the chiral waveguide.
In turn, the dynamics of the ladder operators is governed by 
\begin{equation}
i \partial_t \sigma^-_n(t) =  \sum_{m=1}^N A_{nm}\,\sigma^-_m(t) + \sum_{m=1}^N \sqrt{2\gamma_m}\, c_{mn}(t) \, a_\text{in}(t),
\label{eq:sigma-minus}
\end{equation}
with the matrix $A_{nm}:=\Delta_n\delta_{nm}- i \sqrt{\gamma_n\gamma_m}$.

\section{Scattering matrix}
\label{sec:scattering}

The scattering matrix is defined as the operator that connects states in the asymptotic past with states in the asymptotic future, situations when the photons are not interacting with the scatterer. If $U_I$ is the evolution operator in the interaction picture, the scattering matrix is defined as
 $S^\text{c}= U_I (t_1\to \infty, t_0 \to -\infty)$, where the superscript ``c'' refers to the chiral case. 
  
One of the advantages of the input-output formalism is that it directly provides the connection between those asymptotic states. In what follows, we make use of that connection to relate the scattering matrix elements to the coherences and the excited state population of the scatterer.

\subsection{Single-photon scattering}
In \cite{fan10} the relation\ between $S^\text{c}$ and the input-output theory has been established.  
 The amplitude for the transition from an input state with momentum $k$ into an outgoing state with momentum $p$,  $S_{pk}^\text{c}$, is given by the expectation value
\begin{equation}
S_{pk}^\text{c} = \braket{0|a_\text{out}(p)a_\text{in}^\dagger(k)|0},
\end{equation}
$a_\text{out}(p)= 1/\sqrt {2 \pi } \int {\rm d} t \, a_\text{out} (t) {\rm e}^{i p t}$  is the Fourier transform of the output field. Similarly, $a_\text{in}^\dagger(k)$ is the Fourier transform of the input field $a_\text{in}^\dagger(t)$.

Equation \eqref{eq:in_out} gives
\begin{equation}\label{eq:S1}
S_{pk}^\text{c} = \delta(p-k) - i \sum_{n=1}^N \sqrt{2\gamma_n}
\braket{0|\sigma^-_n(p)|k},
\end{equation}
where 
\begin{equation}
\braket{0|\sigma^-_n(p)|k} :=
\int \frac{e^{ipt}}{\sqrt{2\pi}}
\braket{0|\sigma^-_n(t)|k} \mathrm{d}t,
\end{equation}
and $\ket{k} := a_\text{in}^\dagger(k)\ket{0}$ is the input state with momentum $k$. The dynamics of the matrix elements of $\sigma_n^-(t)$ is obtained by using Eq.\ \eqref{eq:sigma-minus} and $\braket{0|a_\text{in}(t)|k}=e^{-ikt}/\sqrt{2\pi}$:
\begin{equation}
i \partial_t \braket{0|\sigma^-_n(t)|k} =
\sum_{m=1}^N A_{nm}\braket{0|\sigma^-_m(t)|k} + \sqrt{2\gamma_n}
\frac{e^{-ikt}}{\sqrt{2\pi}}.
\end{equation}
This equation can be integrated formally. Introducing the solution in Eq. \ref{eq:S1},
\begin{align}
\label{S}
S_{pk}^\text{c} & = t_k^\text{c} \, \delta(k-p),\\
\label{t1p}
t_k^\text{c} & = 1 - i \sum_{n=1}^N\sqrt{2\gamma_n} \, s^n_{k},\\
\label{eq:s}
s^n_{k} & = \sum_{m=1}^N \sqrt{2\gamma_m} \, [( k - A)^{-1}]_{nm},
\end{align}
where the effect of the occupation of the excited levels in the atom affects the transmission through $s^n_{k} = \braket{0|\sigma^-_n(p)|k}$. 

The limit of a qubit ($N=1$) can be trivially recovered. In this case, $A$ is not a matrix, but just a number and
\begin{align}
s_{k} & = \frac{\sqrt{2\gamma}}{k - \Delta + i\gamma}, \\
t_k^\text{c} & = 1 - i \sqrt{2\gamma} s_{k} = \frac{k-\Delta - i\gamma}{k-\Delta + i\gamma}.
\end{align}

As mentioned, the scattering coefficients in the non-chiral case can be obtained from the chiral ones. The non-chiral transmission coefficient is $t_k=(t_k^\text{c}+1)/2$, while the  reflection coefficient is $r_k=t_k-1$ \cite{fan10}. 
It is essential that the decay rates $\gamma_n$ used in previous expressions are those of the non-chiral waveguide. This point deserves clarification: in terms of the microscopic parameters in a real system, the decay rates in a non-chiral waveguide are $\gamma_n^{non-c} = 2 \pi g_n^2$, where the factor of 2 appears because the excitation can couple to two different photon branches (left and right). A chiral waveguide supports only one-photon branch and  $\gamma_n^c = \pi g_n^2$. However, in the calculation of the scattering matrix in the non-chiral case (characterized for a set of  $\left\{g_n\right\}$) we have used an {\it auxiliary} chiral system where coupling occurs only in one channel (the symmetric channel), with an effective coupling $\sqrt{2} g_n$. So, in this auxiliary chiral system the decay rates are $ (\sqrt{2} g_n)^2\pi$, which coincide with those in the real non-chiral case.

With this,  we obtain for the one-photon scattering matrix of the $V^{(N)}$-atom:
\begin{equation}\label{eq:tnc}
t_k = 1-i\sum_{n=1}^N \sqrt{\frac{\gamma_n}{2}}s^n_{k}.
\end{equation}
For $N=2$, these results coincide with those already published \cite{Witthaut2010,Pletyukhov2012}. 

\subsection{Two-photon scattering}

\subsubsection{Chiral Scattering matrix for arbitrary $N$}
Using the same ideas,  we can also compute the two-photon chiral scattering matrix
\begin{equation}\label{eq:S2}
S_{p_1p_2k_1k_2}^\text{c} = \braket{0|a_\text{out}(p_1)a_\text{out}(p_2)a_\text{in}^\dagger(k_1) a_\text{in}^\dagger(k_2)|0}.
\end{equation}

By introducing the identity $\int a_\text{in}^\dagger(k)\ket{0}\bra{0}a_\text{in}(k) \mathrm{d}k$ between $a_\text{out}(p_1)$ and $a_\text{out}(p_2)$, and following \cite{fan10}, we obtain
\begin{align}\label{eq:Sprev}
S_{p_1p_2k_1k_2}^\text{c} &= t_{p_1}^\text{c}\delta(p_1-k_1)\delta(p_2-k_2)+\nonumber
\\
&+t_{p_1}^\text{c}\delta(p_1-k_2)\delta(p_2-k_1)\nonumber\\
&-it_{p_1}^\text{c}\sum_{n=1}^N\sqrt{2\gamma_n}\braket{p_1|\sigma^-_n(p_2)|k_1k_2}.
\end{align}
 The computation of $\braket{p_1|\sigma_n^-(p_2)|k_1k_2}$ requires some algebraic manipulations and is described in Appendix \ref{app:2photon}. Here we present the final result. The two-photon $S^\text{c}$-matrix is the sum of a linear contribution (product of $t_k^\text{c}$ coefficients, given by \eqref{t1p}) and a nonlinear one, 
\begin{align}
\label{S2p}
S_{p_1p_2k_1k_2}^\text{c} &=t_{p_1}^\text{c}t_{p_2}^\text{c}[\delta(p_1-k_1)\delta(p_2-k_2)
+ (k_1\leftrightarrow k_2)] \nonumber\\
&+ iT_{p_1p_2k_1k_2}^\text{c}\delta(p_1+p_2-k_1-k_2),
\end{align}
The nonlinear term $T^\text{c}$ is responsible for the fluorescence spectrum where the individual energy of each photon is not conserved but the total energy is. It reads:
\begin{align}\label{eq:T}
T_{p_1p_2k_1k_2}^\text{c} &=  \frac{t_{p_1}^\text{c} }{2\pi} \sum_{n=1}^N \sqrt{2\gamma_n}s^n_{p_2}
\sum_{m=1}^N (s^m_{p_1})^*(s^m_{k_1}+s^m_{k_2})
\nonumber\\
&+ \frac{t_{p_1}^\text{c}}{2\pi}  \sum_{n=1}^N  s^n_{p_2}(s^n_{k_1}+s^n_{k_2})\sum_{m=1}^N \sqrt{2\gamma_m}
(s^m_{p_1})^*.
\end{align}
This particular expression will be useful later on. However, it is not evident that it is symmetric under the exchange $p_{1} \leftrightarrow p_{2}$ or $k_{1} \leftrightarrow k_{2}$, as it should be. After some manipulations, described in Appendix \ref{app:2photon}, we arrive to an expression where these exchange symmetries are clearly visible:
\begin{align}\label{eq:simplified-T}
&T_{p_1p_2k_1k_2}^\text{c} = \frac{2}{\pi} \frac{1}{(1+i\alpha_{p_1})(1+i\alpha_{p_2})}\times\\
&\times\left(
\frac{\alpha_{p_2}\beta_{p_1 k_1}+\alpha_{p_1}\beta_{p_2 k_1}}{1+i\alpha_{k_1}} + \frac{\alpha_{p_2}\beta_{p_1 k_2}+\alpha_{p_1}\beta_{p_2 k_2}}{1+i\alpha_{k_2}}\right),\nonumber
\end{align}
with
\begin{align}
\alpha_{k} = \sum_{n=1}^{N}\gamma_{n} \frac{1}{k-\Delta_{n}} ,\;
\beta_{kp} = \sum_{n=1}^{N}\gamma_{n}\frac{1}{k-\Delta_{n}}\frac{1}{p-\Delta_{n}}.
\label{eq:alphabeta}
\end{align}

As a check, notice that Eq. \eqref{S2p} satisfies the general structure that the two-photon scattering matrix should have according to the Cluster Decomposition Principle \cite{weinberg1995,xu13}: a term that indicates conservation of the energy of the individual photons (containing two delta functions) and another term that only conserves total energy (the term with a single delta function).  Also, the result previously obtained for $N=1$ \cite{fan10}, $T_{p_1p_2k_1k_2}^\text{c}= (\sqrt{2\gamma}/\pi) s_{p_2}s_{p_1}(s_{k_1}+s_{k_2})$, is recovered by our calculation.

\subsubsection{Chiral Scattering matrix for $N=2$}
Even though the previous expressions must be computed numerically in general, the case of a $V$-atom $(N=2)$ admits a simple analytical expression with two contributions, $T_{p_1p_2k_1k_2}^\text{c} = T_{p_1p_2k_1k_2}^\text{c1}+ T_{p_1p_2k_1k_2}^\text{c2}$, given by
\begin{eqnarray}
T_{p_1p_2 k_1k_2}^\text{c1} &=& \frac{\sqrt{2}}{\pi} \sum_{n=1}^{2} \sqrt{\gamma_n} \, s^{n}_{p_1} s^{n}_{p_2} \left( s^{n}_{k_1}  + s^{n}_{k_2} \right),\label{eq:tv2a}
\\ 
T_{p_1p_2 k_1k_2}^\text{c2} &=& \frac{1}{\sqrt{2} \pi}  \sum_{n=1}^{2}  s^{n}_{p_1} s^{\bar{n}}_{p_2}   \sum_{m=1}^{2} \sqrt{\gamma_{\bar{m}}} \left( s^{m}_{k_1} + s^{m}_{k_2} \right),
\label{eq:tv2b}
\end{eqnarray}
where we have defined $\bar{n} = 2$ if $n=1$, and $\bar{n} = 1$ if $n=2$, and
\begin{equation}
s_k^{n} = \sqrt{2 \gamma_n} \frac{k-\Delta_{\bar{n}}}{(k-\Delta_1+i\gamma_1)(k-\Delta_2+i\gamma_2)+\gamma_1 \gamma_2}
\end{equation}

For completeness, let us recall that the two-photon scattering matrix for the case of 2 collocated 2LS can be written in a similar way, with the same $T_{p_1p_2 k_1k_2}^\text{c1}$ but with a expression for $T_{p_1p_2 k_1k_2}^\text{c2}$ given by \cite{Rephaeli2011}:
\begin{eqnarray}
T_{p_1p_2 k_1k_2}^\text{c2} &=&  - \frac{i}{\sqrt{2} \pi}  \sum_{n=1}^{2}  s^{n}_{p_1} s^{n}_{p_2}   \sum_{m=1}^{2} \sqrt{\gamma_{\bar{m}}} \left( s^{m}_{k_1} + s^{m}_{k_2}\right) \times \nonumber \\
& &\frac{\sqrt{\gamma_1 \gamma_2}}{k_1 + k_2 - \Delta_1 - \Delta_2 + i \gamma_1+ i\gamma_2}.
\label{eq:t2ls}
\end{eqnarray}

\subsubsection{Scattering matrix in the non-chiral case}

All formulas above have been derived for the chiral case, in which there is only one family of propagating photons with positive momenta, $k>0$. The result for a non-chiral medium with left- and right-moving photons can be obtained from the chiral one as \cite{fan10}:  
\begin{align} 
T_{p_1 p_2 k_1 k_2}=\frac{1}{4}T_{|p_1| |p_2| k_1k_2}^\text{c},
\end{align}
with the prescription that when computing the non-chiral $T$ the decay rates used in the expression for $T^\text{c}$ should be those of the non-chiral system (as discussed at the end of Section III A).

\subsubsection{Flat band}
\label{sec:flat}

The formulas above are also valid in the limit $N>>1$. This situation may describe, for instance, an ensemble of $N$ ultracold atoms which suffer from Rydberg blockade \cite{Jaksch2000,Lukin2001,Saffman10}, where only one of the atoms may absorb a photon at a given time and all other excitations are suppressed [cf. Fig.\ \ref{fig:setup}b]. In this situation, the scattering formulas provide a very compact way of estimating the coupling strength to the ensemble. In particular, it is well known that in the symmetric limit, in which there is no significant inhomogeneous broadening (thus all $\Delta_n \approx \Delta$) and all spontaneous emission rate are very approximately equal ($\gamma_n \approx \gamma$), the system behaves like a ``fat" two-level system, with a bosonic enhancement of the spontaneous emission rate. This fundamental result is recovered from Eq.  \eqref{eq:simplified-T}, which automatically gives $\alpha_k^{(N)} = N \alpha_k^{(1)}$, i.e., $\gamma^{(N)}=N \gamma$.


\section{Discussion}
\label{sec:discussion}

The following subsections deal with various applications of the scattering formulas in the non-chiral case, where $g_{-n}= g_{+n}$ in \eqref{H}. We begin by comparing a $V$-atom with two collocated 2LS. We show that, while single-photon scattering cannot distinguish both experimental setups, remarkable differences appear in their two-photon scattering. We also show that a $V^{(N)}$-level scheme exhibits CRIT in the two-photon scattering spectrum, i.e., for some values of the incoming photon energies the two-photon transmission is perfect and all nonlinear phenomena cancel out due to destructive interference. 

\subsection{Two-photon fluorescence}
\label{sec:2ph-fluorescence}
\begin{figure}[tbh!]
\includegraphics[width=\linewidth]{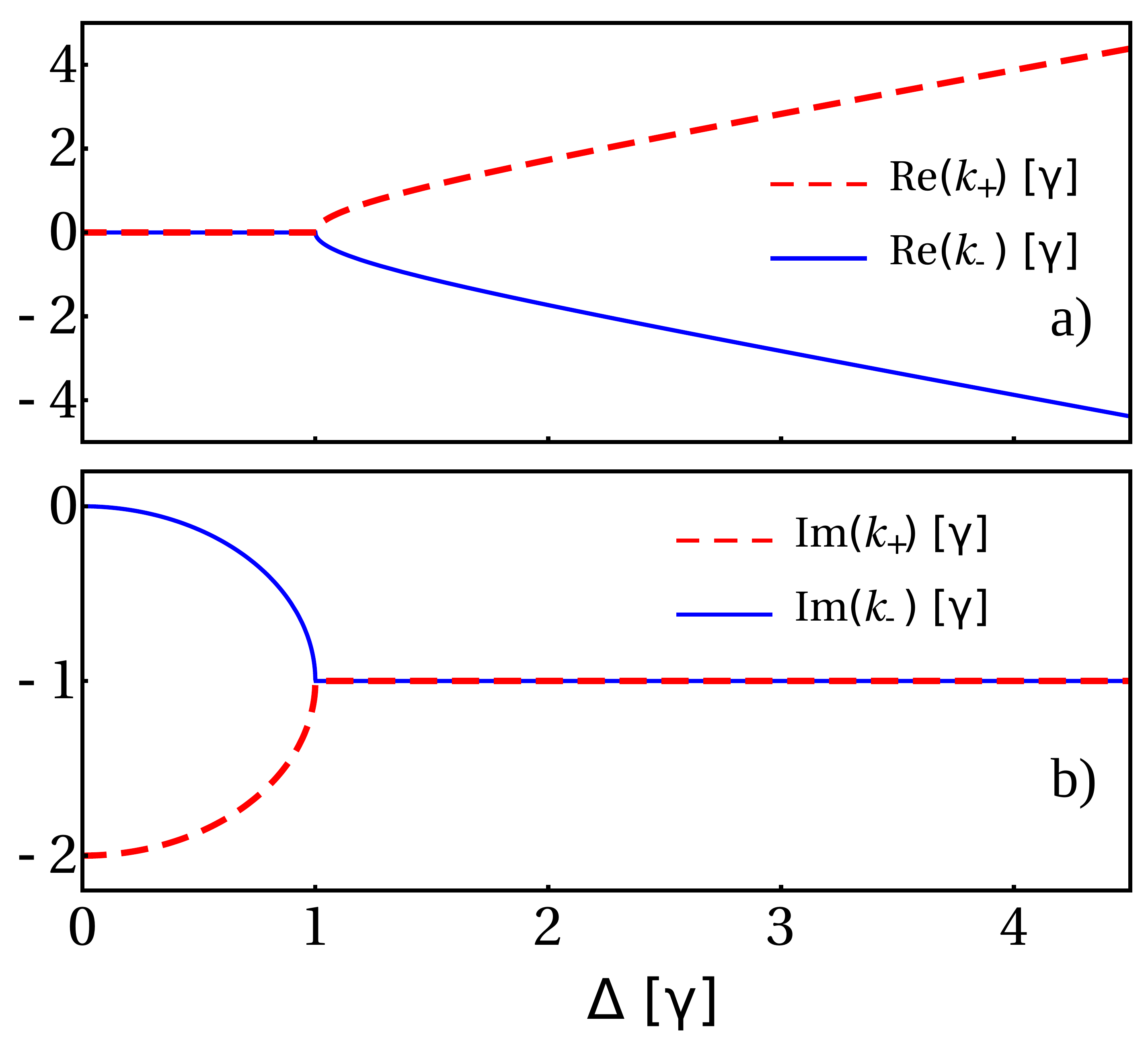}
\caption{(Color Online).  Real and Imaginary part of the single-particle poles of the scattering matrix as a function of $\Delta\equiv \Delta_1=-\Delta_2$. These results apply to both scattering by a $V$-atom and by two collocated 2LS. We have assumed that $\gamma_1=\gamma_2=\gamma$.}
\label{fig:Poles}
\end{figure}

\begin{figure}[tbh!]
\includegraphics[width=\linewidth]{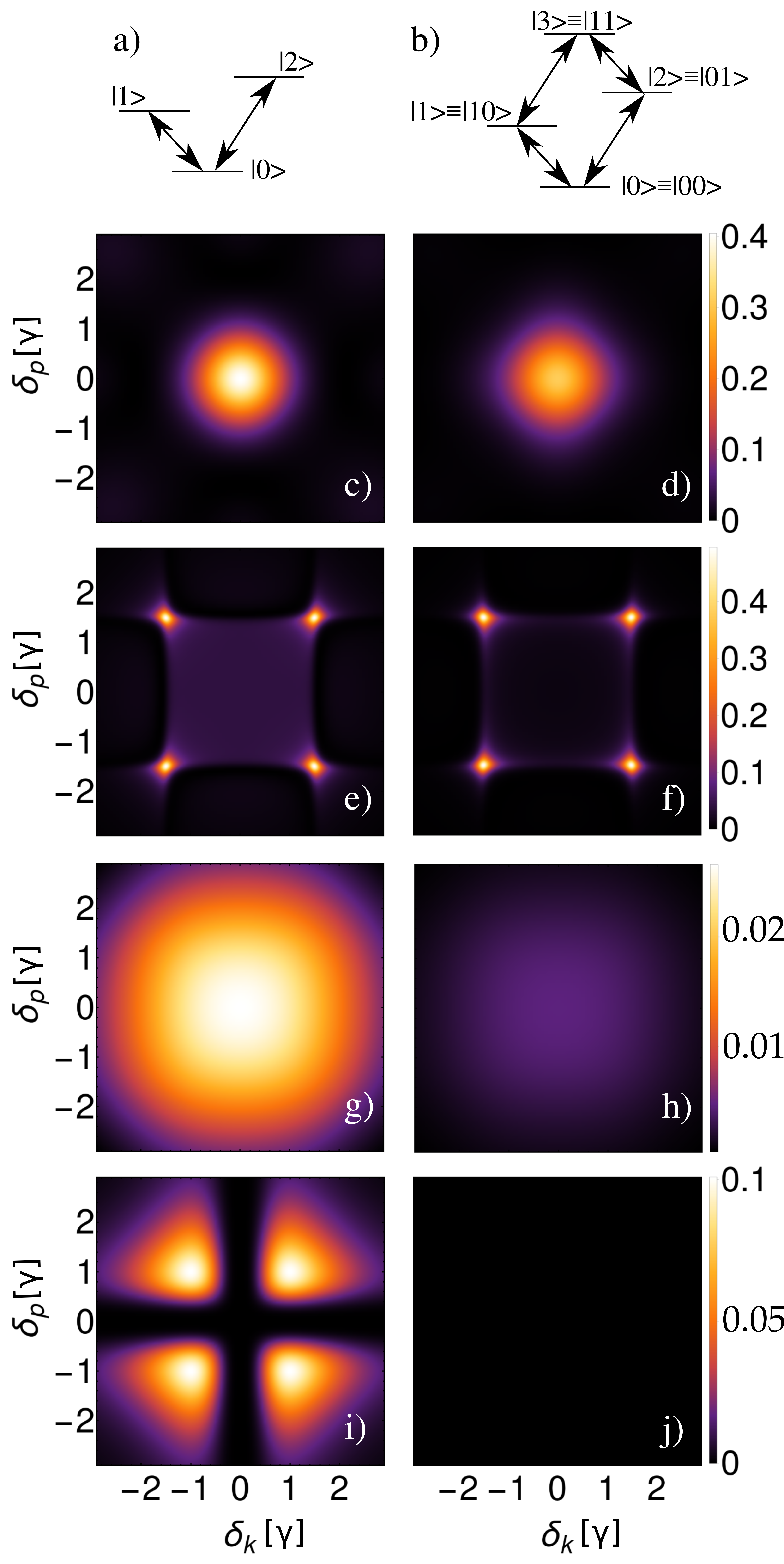}
\caption{(Color Online).  Intensity for resonance fluorescence: $|T_{p_1 p_2 k_1 k_2}|^2$ in units of $1/\gamma^2$ for a $V$-atom (left panels) and two collocated 2LS (right panels), see the level structures drawn on the top. In all cases $\gamma_1=\gamma_2=\gamma$ and $\Delta_1=-\Delta_2 \equiv \Delta$. We define $\delta E \equiv k_1+k_2$. Top panels a) and b) schematically show the level structures.  In panels c) and d), $\Delta= 1.5\gamma$ and $\delta E=3\gamma$. In panels e) and f), $\Delta = 0.5 \gamma$ and $\delta E=3\gamma$. In panels g)  and h), $\Delta = 0$ and $\delta E=3\gamma$,  and in panels i) and j) $\Delta= \gamma $ and $\delta E=0$.}
\label{fig:F}
\end{figure}

We use the previous expressions in order to analyze how much information can be extracted from a two-photon spectroscopy.  For this, we concentrate on the $N=2$ case (a $V-$atom) and, for simplicity, consider that both excitations have the same spontaneous emission rate, $\gamma_1 = \gamma_2= \gamma=1$, which thus sets the unit of energy.  Without loss of generality, we assume $\Delta\equiv \Delta_1=-\Delta_2$, which means that we have chosen the zero of energy to be located at $(\Delta_1+\Delta_2)/2$.

Let us recall that the single-photon transmission, see Eq. \eqref{eq:tnc}, vanishes when the photon energy matches an excitation energy in the scatterer \cite{shen05a,shen05b,Nori2008a}. 
A two-photon transmission spectroscopy may provide extra information, beyond revealing the excitation energies. If any, this effect should be contained in the nonlinear part of the scattering matrix, $T_{p_1 p_2 k_1 k_2}$. In order to analyze the two-photon scattering by a  $V$-atom it is convenient to compare it with that by two collocated 2LS,  which has already been discussed in Ref. \cite{Rephaeli2011}. Notice that the {\it single}-photon scattering is identical in these two cases because they present the same single excitation manifold (see level structure in Fig. \ref{fig:F}, panels a and b).

The analysis of the results is facilitated by the knowledge of  the poles of $T_{p_1 p_2 k_1 k_2}$. For both the $V-$atom and the two collocated 2LS,  $T_{p_1 p_2 k_1 k_2}$ presents poles  at the same spectral positions as the single-particle scattering amplitudes $s_k^n$,  Eqs. \eqref{eq:tv2a}, \eqref{eq:tv2b} and \eqref{eq:t2ls}. There are two kinds of single-particle poles, corresponding to scattering through the states $|\pm \rangle = \frac{1}{\sqrt{2}} (|1 \rangle \pm |2\rangle)$ (see panels a) and b) in Figure \ref{fig:F}), which form a basis spanning the two single-excitations of the scatterers. The spectral position of these poles as a function of $\Delta $ is shown in Fig. \ref{fig:Poles}. Two regimes can be differentiated: when $\Delta>\gamma$, the two excitations essentially behave as independent ones. They are spectrally located at approximately $\pm \Delta$ and present an amplitude decay rate that coincides with the ``bare" rate, $\gamma$. For $\Delta<\gamma$, the two excitations hybridize leading to a super-radiant and a sub-radiant state, both of them spectrally located at the average frequency of the two bare excitations.
Additionally, the scattering by two 2LS give rise to a ``collective" two-photon pole at $k_1+k_2=\Delta_1+\Delta_2 + i (\gamma_1+\gamma_2) $ \cite{Rephaeli2011}, which is not present in the case of scattering by a $V-$atom. 

A representative set of results is shown in Fig. \ref{fig:F}, where we plot $|T_{p_1 p_2 k_1 k_2}|^2$ as a function of  both $\delta_k = (k_1 - k_2)/2$ and $\delta_p = (p_1- p_2)/2$.
Each panel considers different total frequencies of the incident photons, $\delta E := k_1+k_2$, and excitation energies, $\pm \Delta$. Left panels show the results for the $V$-atom, while the right panels render the ones for the two collocated 2LS.  In all panels, the 4-fold rotational symmetry of $|T_{p_1 p_2 k_1 k_2}|^2$ arises from a combination of the indistinguishability of the photons (which makes  $S_{p_1 p_2 k_1 k_2}$ invariant under the interchange $k_1 \leftrightarrow k_2$ or $p_1 \leftrightarrow p_2$) and time-reversal symmetry (which makes $S_{p_1 p_2 k_1 k_2} = S_{k_1 k_2 p_1 p_2}$ \cite{taylor}). 
 
Let us first discuss the case where the two incoming photons cannot be in resonance with both single-particle states, this is, when $ |k_1+k_2 - \Delta_1 - \Delta_2| > 0$. An analysis of this case shows that the intensity for fluorescence $|T_{p_1 p_2 k_1 k_2}|^2$ is maximum when {\it one} of the incoming photons and {\it one} of the outgoing photons are resonant with one of the single-photon transitions. Depending on the difference between the bare excitation energies, we can differentiate two situations.
The first one is when the excitation levels are essentially uncoupled: $\Delta > \gamma$. This instance is represented in panels c) and d) of Fig. \ref{fig:F}. Resonances occur at photon energies $\approx \pm \Delta$, and decay with a rate $\gamma$ (see Fig. \ref{fig:Poles}). In terms of $\delta_k$ and $\delta_p$ this implies that $|T_{p_1 p_2 k_1 k_2}|^2$ is maximum for $\delta_p = \pm \delta_k = \pm (\delta E - 2 \Delta)/2$, (which in the case represented in the figure implies $\delta_p = \delta_k = 0$). 
The second situation appears when the excitation energies strongly couple, i.e., when  $ \Delta< \gamma$. Now, both single-photon transitions occur at zero energy, and thus the two-photon resonance appears at  $\delta_p = \pm \delta_k = \pm \delta E/2$. One of the transitions is super-radiant, while the other one is sub-radiant and shows up as a narrow peak in the intensity for resonance fluorescence (panels e) and f) of Fig. \ref{fig:F}). As $\Delta \rightarrow 0$, the spectral width of the sub-radiant state narrows but, additionally, its coupling to the incoming photons vanishes when $\gamma_1=\gamma_2$. In the limit $\Delta=0$ (shown in panels g) and h) of Fig. \ref{fig:F}) $|-\rangle$ is a dark state and the $V$-atom is exactly mapped into a single 2LS, with a single excited state given by $| +\rangle$ and a modified spontaneous emission rate $2\gamma$. The fluorescence is only due to the super-radiant state and, correspondingly, the maximum fluorescence is now much smaller than when the sub-radiant state dominates. The two 2LS are mapped to a three-level atom, with excited states $|+\rangle$ and $|11\rangle$, and cascaded transitions with equal excitation energies.  The existence of the two-photon state in the two 2LS diminishes the photon-photon interaction with respect to that of the $V-$atom.

This analysis shows that in the non-resonant case the difference between the fluorescence of the  $V$-atom and the pair of 2LS is quantitative. The nonlinearity is higher for the $V$-atom, because it is more sensitive to saturation effects than the pair of 2LS. 

A different situation arises when both incoming photons may be in resonance with the two single-photon transitions, i.e., when  $k_1+k_2 = 0$. Then, the two 2LS can simultaneously scatter two photons and the nonlinear contribution to the scattering matrix vanishes \cite{Rephaeli2011} (see panel j). In contrast, the $V$-atom does not present the doubly excited state $|11\rangle$ and the fluorescence cannot be quenched. The intensity of resonance fluorescence is maximum when the energies of each incoming photon equals those of the excitations in the $V$-atom (see panel i).

Notice, however, that fluorescence quenching, $T_{p_1p_2k_1k_2}=0$, also appears in the scattering by the $V$-atom, when $\delta_k=0$ and $\delta_p=0$. We explain this effect in the following subsection.

Lastly, note that we have considered two collocated 2LS which do not interact each other. The presence of dipole-dipole interaction can be straightforwardly taken into account as two interacting 2LS can be mapped to a new pair of independent 2LS with modified energies and coupling constants. Thus, any pair of interacting collocated 2LS has a corresponding $V$-atom with the same effective energies and coupling constants.

%
\subsection{Two-photon CRIT interference}
\label{sec:2photon-eit}
%
%
\begin{figure}
\includegraphics[width=\linewidth]{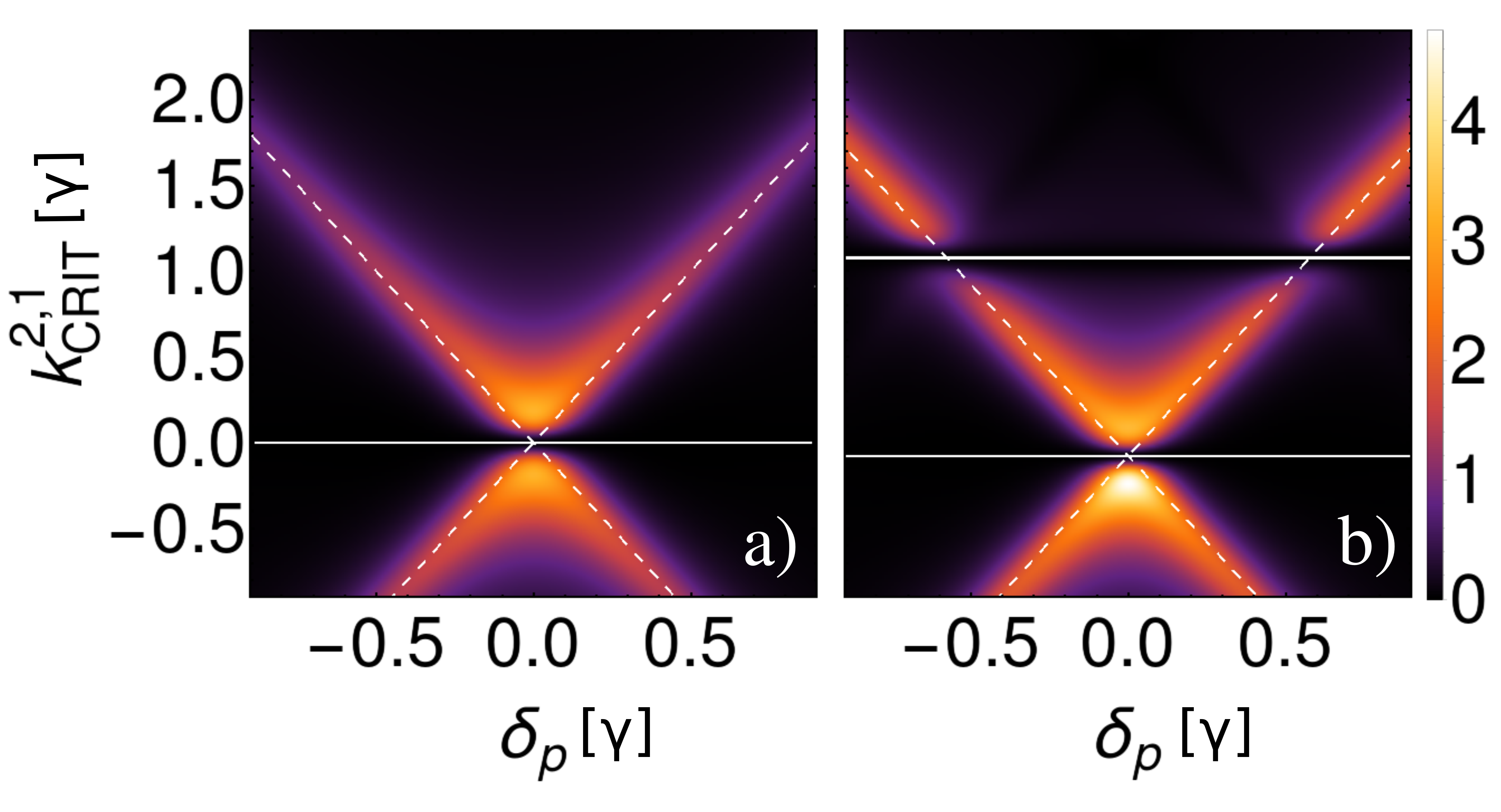}
\caption{(Color Online).  Intensity for resonance fluorescence: $|T_{p_1 p_2 k_1 k_2}|^2$ in units of $1/\gamma^2$, with fixed $k_1=k_\text{CRIT}^{2,1}$, vs $k_2-k_\text{CRIT}^{2,1}$ and $\delta_p$ for $N=2$ (panel a) and $N=3$ (panel b). We fix $\gamma_j=\gamma$ and $\Delta_{j+1}-\Delta_j=\gamma$. The solid white lines represent $k_2=k_\text{CRIT}^{2,1}$ in the left panel, and $k_2=k_\text{CRIT}^{3,1}$ (bottom) and $k_2=k_\text{CRIT}^{3,2}$ (top) in the right panel. The dashed white lines render the condition for the individual conservation of both photon energies, $p_1=k_1$ and $p_2=k_2$, or vice-versa.}
\label{fig:CRIT}
\end{figure}

%
The coupling of a single propagating photon to two or more resonant transitions can produce situations where the transmission is perfect, a phenomenon denoted as Coupled-Resonator-Induced Transparency
 \cite{Smith2004}. 
According to  Eq. \eqref{eq:tnc}, perfect single-photon transmission occurs whenever the input frequency matches the condition:
\begin{equation}
\sum_{n=1}^N \sqrt{\gamma_j}s^n_{k}=0
\label{eq:wEIT}
\end{equation}
This condition can be recast into a $(N-1)$-degree polynomial in $k$, with $N-1$ roots, $k_\text{CRIT}^{N,n}$ ($n=1,\dots,N-1$).
For the $N=2$ case, the condition for transparency is:
\begin{equation}
\label{eq:vEIT}
k_\text{CRIT}^{2,1} = \frac{\gamma_2\Delta_1+\gamma_1\Delta_2}{\gamma_1+\gamma_2}.
\end{equation}

The computed two-photon scattering matrix allows the study of the conditions which lead to the vanishing of the nonlinear term $T_{p_1 p_2 k_1 k_2}$, which is responsible for both fluorescence and photon-photon interaction. Previous studies have found fluorescence quenching for the two-photon power spectrum in a $V-$atom ($N=2$) illuminated with {\it classical} light \cite{Zhou1996}, and also in the case of a driven $\lambda$-system when the incoming photons satisfy the single-photon CRIT condition \cite{Fang2016}.  

For the case of a $V^{(N)}$-atom, it is easy to show that $T_{p_1 p_2 k_1 k_2}=0$ whenever each incoming photon satisfies a single-photon CRIT condition. For this, we first consider that the {\it outgoing} photons satisfy $p_1=k_\text{CRIT}^{N,j}$ and $p_2=k_\text{CRIT}^{N,l}$. Then, introducing the CRIT condition Eq. (\ref{eq:wEIT}) in Eq. (\ref{eq:T}), we obtain  $T_{p_1p_2k_1k_2}=0$, for any pair of incoming photons and that particular channel for outgoing photons. As time-reversal symmetry implies $T_{p_1p_2k_1k_2}= T_{k_1k_2p_1p_2}$, we obtain that $T_{p_1p_2k_1k_2}=0$ whenever the incoming photons satisfy the single-photon CRIT conditions, for any value of the outgoing photon energies. Notice that this derivation also applies to the driven $\lambda$-atom as, in the system eigenbasis $|\pm \rangle $, it can be mapped to a $V$-atom. 
This fluorescence quenching is  shown in Fig. \ref{fig:CRIT}, where we represent $|T_{p_1 p_2 k_1 k_2}|^2$,  for both $N=2$ and $N=3$, when one input photon frequency is taken at $k_\text{CRIT}^{N,1}$, while the frequency of the other incoming photon frequency varies. 
 We already saw this effect in Fig. \ref{fig:F}, panel i). In that case, the CRIT condition for the input photons is fulfilled for  $k_1=k_2$, so $T_{p_1p_2k_1k_2}=0$ for $\delta_k=0$. In the same way, $T_{p_1p_2k_1k_2}$ also vanishes when the output energies satisfy $\delta_p=0$.

If one of the photons is not at a CRIT condition, photon-photon interactions emerge, being maximal when the individual energies of the outgoing photons coincide with those of the incoming ones  (dashed lines in Fig. \ref{fig:CRIT}), as explained in the previous subsection. 

\begin{figure}
\includegraphics[width=\linewidth]{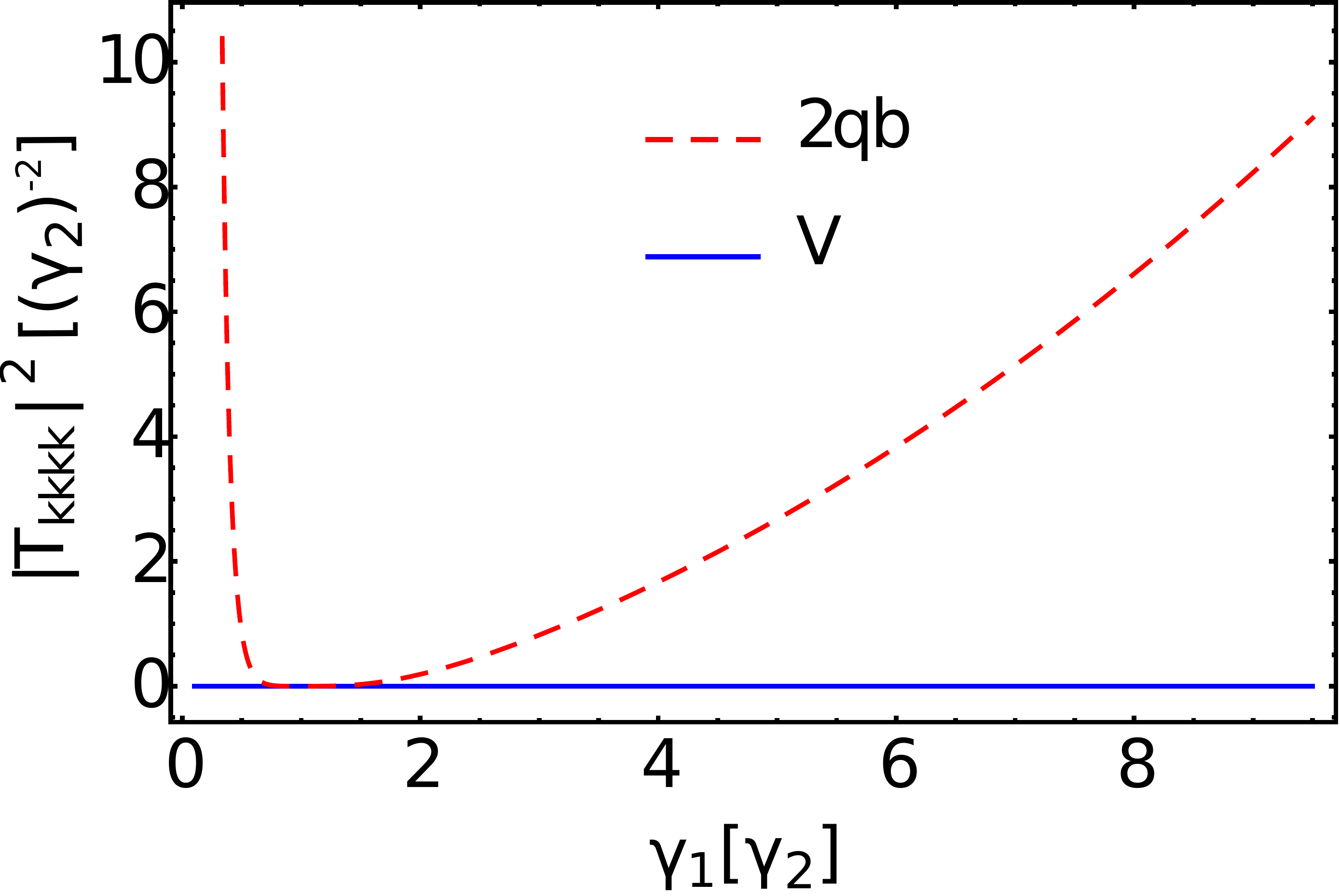}
\caption{(Color Online).  Intensity for resonance fluorescence: $|T_{k k k k}|^2$, with fixed $k =k_\text{CRIT}^{2,1}$, as a function of $\gamma_1$, for both a $V-$atom (blue, solid curve) and two 2LS (red, dashed). We have taken $\Delta=\gamma_2$.  Notice that fluorescence quenching only occurs at $\gamma_1=\gamma_2$ for the two 2LS, but it always vanishes at the two-photon CRIT condition for the $V-$atom. }
\label{fig:Quenching}
\end{figure}

Notice that the statement that fluorescence is quenched in a two-photon scattering process whenever both incident photons satisfy a CRIT condition, which occurs for a  $V^{(N)}$-atom, does not necessarily apply to all possible scatterers. A counterexample is the case of two collocated 2LS. There, fluorescence quenching occurs  when the total energy of the incoming photons is equal to the sum of the excitation energies ($k_1+k_2=\Delta_1+\Delta_2$), but only when both 2LS couple equally to the waveguide ($\gamma_1=\gamma_2$)\cite{Rephaeli2011}. As shown in Fig. \ref{fig:Quenching}, if these couplings are unequal, the two 2LS present a non-vanishing resonance fluorescence when the incoming photons are at individual CRIT conditions, $k=k_\text{CRIT}^{2,1}$. The chosen output frequencies are also $k$, but this is irrelevant, as other choices would only change the intensity of the fluorescence, but not the overall dependence on $\gamma_1/\gamma_2$. In contrast, in the $V$ case, the fluorescence is not generally quenched when the total energy of the incoming photons is equal to the sum of the excitation energies. But, when each of the two incoming (or outgoing) photons is in single-photon CRIT conditions, both of them are transmitted with unit amplitude and the fluorescence is quenched, even for dissimilar couplings of the excitations to the waveguide (see Fig. \ref{fig:Quenching}). 

\section{Summary and outlook}
\label{sec:summary}

In this work we have developed the single- and two-photon scattering theory for a $V^{(N)}$-level scatterer coupled to either a chiral or a non-chiral waveguide.  We have highlighted that a two-photon spectroscopy can characterize different level structures that would be indistinguishable in a single-photon experiment.  Besides, we have introduced the concept of 
two-photon CRIT.  We have shown that in the $V^{(N)}$-atom structure the two-photon resonance fluorescence is completely quenched when each photon is at single-photon CRIT condition. This can be understood as the quantum version for the phenomenon of fluorescence quenching which occurs when driving a $V-$atom with classical light\ \cite{Zhou1996}. These effects can be seen in the laboratory with state-of-the-art technologies in systems like atoms with a $V-$level structure, or collections of Rydberg atoms where a blockade mechanism prevents simultaneous multi-excitation. 

\begin{acknowledgments}
This work has been supported by Spanish Mineco projects  FIS2012-33022, FIS2014-55867-P and  MAT2014-53432-C5-1-R, CAM Research Network QUITEMAD+, the Gobierno
de Arag\'on (FENOL group) and EU project PROMISCE.
\end{acknowledgments}

\appendix

\section{Derivation of the two-photon chiral scattering matrix}
\label{app:2photon}

Our derivation follows along the lines described in Ref.\ \cite{fan10}. Here, we sketch the major deviations from that reference. The crucial element in the scattering matrix is the Fourier transform of the off-diagonal element of the scatterer between different input and output states, Eq. \eqref{eq:Sprev}:
\begin{align}
&\braket{p_1|\sigma^-_n(p_2)|k_1k_2} :=\\
&\quad\int\frac{e^{ip_2t}}{\sqrt{2\pi}}\braket{0|a_\text{in}(p_1)\sigma^-_n(t)a_\text{in}^\dagger(k_1) a_\text{in}^\dagger(k_2)|0}\mathrm{d}t.\nonumber
\end{align}
The equations for the integrand can be found from Eq.\ \eqref{eq:sigma-minus}
\begin{align}
i \partial_t \braket{p_1|\sigma^-_n(t)|k_1k_2} &=
\sum_{m=1}^N A_{nm}\braket{p_1|\sigma^-_m(t)|k_1k_2}\\
& + \sum_{m=1}^N\sqrt{2\gamma_m}\braket{p_1|c_{mn}(t)a_\text{in}(t)|k_1k_2}.\nonumber
\end{align}
The second term in this equation can be simplified as a transition amplitude between single-photon states
\begin{align}
\braket{p_1|c_{mn}(t)a_\text{in}(t)|k_1k_2} &=
\braket{p_1|c_{mn}(t)|k_1}\frac{e^{-ik_2t}}{\sqrt{2\pi}}
+ (k_1\leftrightarrow k_2).
\end{align}
We now expand $c_{mn}(t)$
\begin{align}
\braket{p|c_{mn}(t)|k} &=\delta_{mn}\braket{p|1-\sum_l \sigma^+_l(t)\sigma^-_l(t)|k}\\
&-\braket{p|\sigma^+_m(t)\sigma^-_n(t)|k}\nonumber
\end{align}
and use the relation
\begin{align}
\braket{p|\sigma^+_m(t)\sigma^-_n(t)|k}=\frac{e^{i(p-k)t}}{2\pi}
(s^m_{p})^*s^n_{k}.
\end{align}
We define $\mathbf{v}(t)$ as a vector whose entries are $v_n(t)=\braket{p_1|\sigma^-_n(t)|k_1k_2}$.  In terms of these quantities we obtain
\begin{equation}
i \partial_t \mathbf{v}(t) = A \mathbf{v}(t)
+ \mathbf{f}_1 \frac{e^{-ik_1t}}{\sqrt{2\pi}}
+ \mathbf{f}_2 \frac{e^{-ik_2t}}{\sqrt{2\pi}}
+ \mathbf{f}_{12} \frac{e^{-ipt}}{\sqrt{2\pi}},
\label{eq:v}
\end{equation}
where $p=p_1-k_1-k_2$ and we have defined the auxiliary vectors
\begin{align}
f_{1,n} =& \sqrt{2\gamma_n} \, \delta(p_1-k_2),\\
f_{2,n} =& \sqrt{2\gamma_n} \, \delta(p_1-k_1),\\
f_{12,n} =& 
-\sqrt{2\gamma_n}\sum_{m=1}^N \frac{1}{2\pi}(s^m_{p_1})^*(s^m_{k_1}+s^m_{k_2})\\
&
-\sum_{m=1}^N \sqrt{2\gamma_m}\frac{1}{2\pi}(s^m_{p_1})^*(s^n_{k_1}+s^n_{k_2}).\nonumber
\end{align}
Equation\ (\ref{eq:v}) can be readily integrated. Taking the Fourier transform in the time variable, we find,
\begin{equation}
\mathbf{v}(p_2) = \mathbf{v}_1 
+\mathbf{v}_2 
+\mathbf{v}_{12},
\end{equation}
with the vectors
\begin{align}
\mathbf{v}_{1} =&   (k_1-A)^{-1} \, \mathbf{f}_1 \, \delta(p_2-k_1),\\
\mathbf{v}_{2} =&   (k_2-A)^{-1} \, \mathbf{f}_2 \, \delta(p_2-k_2),\\
\mathbf{v}_{12} =&  (p_2-A)^{-1} \, \mathbf{f}_{12} \, \delta(p_2-p).
\end{align}
Introducing this relations into Eq. \eqref{eq:Sprev}, and applying\ \eqref{t1p}, we get the expression \eqref{S2p} for $S_{p_1p_2k_1k_2}^\text{c}$, with $T_{p_1p_2k_1k_2}^\text{c}$ given by\ \eqref{eq:T}.

The problem with the previous standard derivation and the final formula\ \eqref{eq:T} is that it hides the exchange symmetry between outgoing bosons $p_1$ and $p_2$. To recover this symmetry we have to realize that it is possible to manipulate the expression for $\mathbf{s}_k$ to simplify all the sums. We begin by writing the innards of $\mathbf{s}_k$ explicitly
\begin{align}
(k-A) &= E_k^{1/2}(1+i \mathbf{w} \mathbf{w}^\dagger)E_k^{1/2},
\end{align}
in terms of a diagonal matrix $E_{k,nm} = (k - \Delta_n)\delta_{nm}$ and the unnormalized vector $w_n = \sqrt{\gamma_n}$. Introducing the factor
\begin{equation}
\alpha_k = \mathbf{w}^\dagger E_k^{-1} \mathbf{w},
\end{equation}
we arrive at the expression
\begin{equation}
(k-A)^{-1} = E_k^{-1}\left[1 - \frac{i}{1+\alpha_k}\mathbf{w} (E_k^{-1}\mathbf{w})^\dagger\right].
\end{equation}
We can use this simplification to write
\begin{align}
\mathbf{s}_k = \frac{\sqrt{2}}{1+i\alpha_k} E_k^{-1}\mathbf{w},\;t^c_k &=\frac{1-i\alpha_k}{1+i\alpha_k},
\label{eq:simplified-t}
\end{align}
which shows that the chiral transmission coefficient is just a phase.

We can achieve a similar simplification of the two-photon scattering matrix identifying sums with scalar products
\begin{align}
\sum_l \sqrt{\gamma_n} s^n_{p} &= \mathbf{w}^\dagger \mathbf{s}_p = \frac{\sqrt{2}\alpha_p}{1+i\alpha_k},\\
\sum_n (s^n_{p})^* s^n_{k}&=\mathbf{s} ^\dagger_p \mathbf{s}_k = \frac{2}{(1-i\alpha_p)(1+i\alpha_k)}\beta_{pk}
\end{align}
to first transform Eq.\ \eqref{eq:T}
\begin{align}
T_{p_1p_2k_1k_2}^\text{c} &=  \frac{t_{p_1}^\text{c} }{2\pi}
\sqrt{2}(\mathbf{w}^\dagger \mathbf{s}_{p_2})
[\mathbf{s}_{p_1}^\dagger(\mathbf{s}_{k_1}+\mathbf{s}_{k_2})]
\nonumber\\\
&+\frac{t_{p_1}^\text{c} }{2\pi}
\sqrt{2}(\mathbf{s}_{p_1}^\dagger \mathbf{w})
[\mathbf{s}_{p_2}^T(\mathbf{s}_{k_1}+\mathbf{s}_{k_2})],
\end{align}
and then transform it to the final expression in Eq. \eqref{eq:simplified-T}.

\bibliographystyle{apsrev4-1}
\bibliography{chiral}

\end{document}